\documentclass[a4paper,11pt]{article}

\usepackage[latin1]{inputenc}
\usepackage[OT1]{fontenc} 
\usepackage[french]{babel} 
\usepackage{verbatim}
\usepackage{multicol}
\usepackage{ulem}
\usepackage{times}
\usepackage{graphicx}
\usepackage{moriond,epsfig}

\def\Journal#1#2#3#4{{#1} {\bf #2}, #3 (#4)}

\def\AA{\em Astronomy \& Astrophysic}
\def\AJ{\em Astromical Journal}
\def\APJ{\em Astrophysical Journal}
\def\APH{\em Astro.ph}

\begin{document}
\vspace*{4cm}
\title{DERIVATION OF DISTANCES WITH THE TULLY-FISHER RELATION: THE ANTLIA CLUSTER}

\author{ N. Bonhomme$^{(1)}$ , H.M.Courtois$^{(1)}$ , R.B.Tully$^{(2)}$ }

\address{(1) Institut de Physique Nucleaire de Lyon, 4 rue Enrico Fermi,\\
domaine scientifique de la DOUA 69622 Villeurbanne CEDEX , France\\
(2) Institute for Astronomy, University of Hawaii, 2680 Woodlawn Drive, Honolulu, HI 96822}

\maketitle\abstracts{
The Tully-Fisher relation is a correlation between the luminosity and the HI 21cm line width in spiral galaxies (LLW relation). It is used to derive galaxy distances in the interval 7 to 100 $Mpc$. Closer, the Cepheids, TRGB and Surface Brightness Fluctuation methods give a better accuracy. Further, the SNIa are luminous objects still available for distance measurement purposes, though with a dramatically lower density grid of measurements on the sky. Galaxies in clusters are all at the same distance from the observer. Thus the distance of the cluster derived from a large number of galaxies (N)  has an error  reduced according to $\sqrt{N}$.
However, not all galaxies in a cluster are suitable for the LLW measurement.
The selection criteria we use are explained hereafter; the important point being to avoid Malmquist bias and to not introduce any systematics in the distance measurement.}

\section{Selecting galaxies in Antlia cluster}
The authors have an ongoing large project of measuring 14 clusters for purpose of calibrating the
slope of the LLW relation at large distances. In this proceedings, we present an example using
the Antlia Cluster. The selection steps use the extensive LEDA \cite{patu} database as follows:\\
1.List all galaxies that are possible members of the cluster : based on radial velocities, and coordinates above a given magnitude limit (Mk $\leq$ -21 , 2MASS survey) \cite{jar}.\\
2.Using the morphological type, discriminate between elliptical and spiral galaxies. Only galaxies of type later than $S0$ (not included) have enough HI gas to be reliably detected with radio observations.\\
3.The candidates should have an inclination $(i)$ between 90 and 45 degrees in order to measure the Doppler shift due to the gas rotation in the galaxy and to obtain an accurate optical measurement of the inclination $(i)$ . The galaxy should also not be an interacting or have a confusing nearby galaxy in the radio telescope beam. \\

\section{Photometry}

It has been shown \cite{tu} that the use of shorter exposures as in 2MASS survey can lead to systematics in the distance derivation for the LLW, since the extended part of the galaxy disk are not probed by short exposures.  This is clearly understood by the fact that the radio HI profile is correlated to the outer parts of the disk. Deep exposures (300 sec) in I band are currently obtained at UH2.2m telescope (Hawaii).
The archangel software \cite{sch}  is used to compute the surface brightness and total integrated magnitudes of a galaxy.
Flux calibration is provided by standard Landolt star exposures taken throughout the night.
Critical points in the magnitude determination are the stars masking and the sky level determination since we try to extend the measurements to the farther outskirts of
the galaxy. In the example, the measure of the apparent magnitude of pgc31995 with Archangel \cite{sch}
gives $ m_I = 11.32 \pm 0.03$, the inclination $i = 78^o$ and the apparent magnitude in I band, corrected by extinction of our galaxy $(b)$, inclination of the target galaxy $(i)$ and the redshift effect $(k)$ is $m_I^{b,i,k} = 10.45 \pm 0.04$. 

\section{HI linewidth}

The authors are involved in a large program involving the collection of HI spectra from Arecibo, GBT, Parkes and
Nancay radiotelescopes. 
 The critical points in the linewidth measurement
are:\\
-Clearly defining the beginning and the end of the spectral lines edges.\\
-Measuring the width at a constant physically meaningfull level independent of the line profile shape (single peak, double peak, narrow, high or low).
The literature is abundant with discussions about which way the HI line should be measured. The
authors have been developping a new algorithm to measure the profile at 50\% of the integrated
line rather than 20\% or 50\% of the peaks level. The linewidth measurement gives in our example:
$W_R = 396 \pm 8$ $kms^{-1}$ . 

\begin{figure}
  \centering
  \includegraphics[width=15 cm]{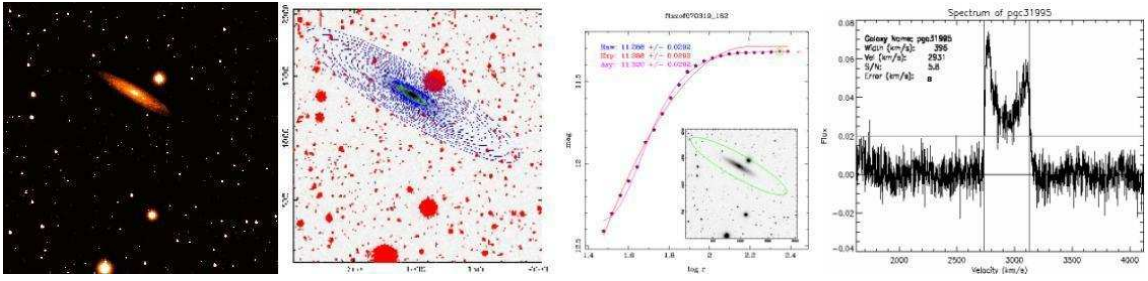}
  \caption{Archangel photometry + HI profile
    \label{fig:arc}}
\end{figure}

\section{Results}
The most recent calibration of the LLW with cepheids \cite{tu} gives: $M_I^{b,i,k} = -21.43 - 8.11 (logW_R^i -2.5)$.
$M_I^{b,i,k}$ is the absolute magnitude in I band and 
$W_R^i$ is the linewidth corrected by the inclination $(i)$ of the target galaxy: $W_R^i = W_R / sin(i)$.
Once the slope of the LLW is calibrated from clusters studies, the LLW can be used to derive distances of field galaxies in the local universe volume up to 10,000 km/s. A distance allows a subtraction of expansion velocity  $H_0d$, from the observed velocity $V_{obs}$  to give a peculiar velocity:  $V_{pec}$ = $V_{obs}-H_od$ . For pgc31995, $M_I^{b,i,k}=-22.30 \pm 0.35$ and then $d = 35.5 \pm 6$ $Mpc$. One realize that an accuracy of better than 17\% is reached for a single galaxy. We might reached 4 or 5\% for a cluster with $\sqrt{N}$ improvement. Thus with $H_0 = 73$ $kms^{-1} Mpc^{-1}$ , $V_{pec} = 357 \pm 21$ $kms^{-1}$.
Finally, the peculiar flows are used as a tracer of the luminous and dark matter gravitational fields in the local universe \cite{tu}.

\end{document}